\documentclass[preprint,prd,showpacs,amsmath,amssymb,amsthm,nofootinbib]{revtex4}
\usepackage[mathscr]{euscript}
\usepackage{bm}
\usepackage{graphicx}
\usepackage{subfigure}
\usepackage{overpic}
\usepackage{multirow}
\usepackage{floatrow}
\usepackage[colorlinks=true,linkcolor=red]{hyperref}
\newfloatcommand{capbtabbox}{table}[][\FBwidth]
\newcommand{\bea}{\begin{eqnarray}}
\newcommand{\eea}{\end{eqnarray}}
\newcommand{\beq}{\begin{equation}}
\newcommand{\eeq}{\end{equation}}

\def\/{\over}

\begin{document}

\title{Production of gravitational waves during preheating  in the Starobinsky inflationary model }

\author{ Guoqiang Jin, Chengjie Fu, Puxun Wu 
 and Hongwei Yu}
\affiliation{Department of Physics and Synergetic Innovation Center for Quantum Effects and Applications, Hunan Normal University, Changsha, Hunan 410081, China 
}

\begin{abstract}
The production of GWs during preheating in the Starobinsky model with a nonminimally coupled auxiliary scalar field  is studied through  the lattice simulation in this paper.  We find that the GW spectrum $\Omega_{\rm gw}$  grows fast with the increase of the absolute value of coupling parameter $\xi$. This is because the resonant bands become broad with the increase of $|\xi|$.  When $\xi<0$, $\Omega_{\rm gw}$ begins to grow once the inflation ends and grows faster than the case of $\xi>0$.  $\Omega_{\rm gw}$  reaches the maximum at  $\xi=-20$ ($\xi=42$ for the case $\xi>0$) and then decreases with slight oscillation. Furthermore we find that  the GWs produced in the era of preheating satisfy the limits from the Planck and next-generation CMB experiments.

\end{abstract}


\maketitle
\section{Introduction}
\label{sec_in}
Inflation, a phase of quasi-exponential expansion in the early universe,  is an elegant idea proposed to resolve most of theoretical problems in the big bang standard cosmology~\cite{Inflation1,Inflation2,Inflation3,Inflation4}. Furthermore, the scalar quantum fluctuations during inflation provide the seed for the formation of large scale structure~\cite{Fluc}. Inflation usually is assumed to be driven by a slow-roll single scalar field, which predicts a scale-invariant  spectrum of scalar  fluctuations on the super-horizon scale. This prediction is consistent with the Cosmic Microwave Background (CMB) observations~\cite{Smoot}. The latest CMB limits the spectral index of the power spectrum  to be $n_s=0.9649\pm 0.0042$ at the $68\%$ confidence level (CL)~\cite{Planck2018}. Except for the  scalar perturbations, there also exists the tensor fluctuations during inflation, which is also  scale-invariant on large scales. It is worthy to note that   tensor perturbations in the scope of general relativity were first and quantitatively correctly calculated in~\cite{Starobinsky1979},  and  the quantitatively correct expressions for both scalar and tensor perturbations were first presented in~\cite{Starobinsky1983} for the Starobinsky model based on modified $f(R)$ gravity~\cite{Inflation1}.
Since the tensor perturbations lead to the B-mode polarization for the CMB photon, its amplitude can be constrained by the CMB observations. The allowed region of the ratio of tensor to scalar fluctuations $r$ is $r_{0.002}<0.10$ at the $95\%$ CL~\cite{Planck2018}. Thus, the combination of $n_s$ and $r$ can discriminate a host of inflationary models. In this regard, it is well-known that the Starobinsky inflationary model is in excellent agreement with the latest CMB observations~\cite{Planck2018}.

After inflation, the inflaton usually will undergo a periodical oscillation around the minimum of its potential. During this oscillation  the inflation field  will decay into some light particles  to thermalize the universe, which is called reheating~\cite{reheating}.
In the first stage of reheating, i.e., preheating, the periodical oscillation of the inflaton may lead to an explosive  production of the inflaton quanta or other light particles coupled to the inflation field through the parametric resonance~\cite{Brandenberger,Kofman94,Bassett2006,Amin2015,Fu2017,Zhu2019}.
When the mode momenta of the inflaton quanta or the light fields  are in the resonance bands, these Fourier modes will grow exponentially. Since only a part of modes have the resonant momenta,  the matter distribution has large and time-dependent inhomogeneities in the position space, which results in that  the matter possesses  substantial quadruple moments, and becomes an effective source of significant gravitational waves (GWs)~\cite{Khlebnikov}. The production of GWs during preheating has been studied widely in~\cite{Bellido2007,Bellido2008,Dufaux2007,Dufaux2009,Figueroa2017,Antusch,Adshead2019}. Different from vacuum fluctuations of tensor perturbations during inflation, the amplitude of the GW spectrum generated during preheating is independent of the energy scale of inflation which only determines the present peak frequency~\cite{Easther, Easther2}. For low energy scale inflationary models, the peak frequency of GWs produced after inflation may well occur in the range which in principle can be detected by future direct detection experiments (like LIGO/VIRGO)~\cite{Antusch,Liu2018, Liu2019, Fu2018}. This opens  a unique observational window for us to  test inflation and  the subsequent dynamics of the very early universe.

It has been found that in the Starobinsky model  the creation of matter can be realized through the decay of scalarons~\cite{Inflation1,Starobinsky, Vilenkin}. However,  the inflaton field in the Starobinsky mode is absent of self-resonance, and in order to realize the preheating process an auxiliary  scalar field $\chi$ coupled non-minimally with the gravity is required~\cite{1999Tsujikawa1}. Recently, the rescattering between the $\chi$ particles and the inflaton condensate during preheating has been studied in~\cite{2019Fu} by using the lattice simulation. The result shows that the rescattering is an efficient mechanism promoting the growth of the $\chi$ field variance, and knocks copious inflaton particles out of the inflaton condensate. Thus, both the auxiliary field and the inflaton field become the effective  gravitational wave sources. However, Ref.~\cite{2019Fu} does not investigate the  production of GWs during preheating, which motivates us to finish the present work.

In this paper, we investigate the production of GWs during preheating in the Starobinsky inflation model. The energy density and spectrum of GWs will be analyzed detailedly by using the lattice simulation. Throughout this paper, we adopt the metric signature (-, +, +, +). Latin indices run from 0 to 3, Greek letters do from 1 to 3, and the Einstein convention is assumed for repeated indices.

\section{basics equations}
\label{sec2}
For the Starobinsky model, when an auxiliary scalar field $\chi$ coupled non-minimally with gravity is considered, the Lagrangian has the form~\cite{1999Tsujikawa1}
  \begin{align}\label{1}
  \mathcal{L}=\sqrt{-{g}}\left[\frac{1}{2\kappa^2}\left(R+\frac{R^2}{6\mu^2}\right)-\frac{1}{2}\xi R \chi^2-\frac{1}{2}(\nabla\chi)^2\right]\;,
  \end{align}
  where $g$ is the determinant of metric tensor $g_{\mu\nu}$, $R$ is the Ricci scalar, $\kappa^{-1}=M_{p}$ with $M_p$ being the reduced Planck mass, $\mu$ is a constant, which is  fixed at $1.3\times10^{-5}M_p$ by the magnitude of the primordial density perturbations \cite{Starobinsky1983, 2007Faulkner}, and $\xi$ is an arbitrary coupling parameter.  In our following analyses, besides the weak coupling,  the case of strong coupling $|\xi|\gg1$ is also considered. The part of the reason is that with the increase of $|\xi|$ ($\xi<0$) the value of $r$ decreases and enters the $68\%$ region allowed by the Planck observations in the  nonminimally coupled inflation model with a self-coupling potential~\cite{Tsujikawa2013}.   Usually, it is  convenient to discuss the cosmic dynamics in the Einstein frame.  By taking a conformal transformation: $\hat g_{\mu\nu}=\Omega^2 g_{\mu\nu}$ with $\Omega^2 $ being $\Omega^2=1-\xi\kappa^2\chi^2+R/(3\mu^2)$, and introducing a scalar field $\phi \equiv \sqrt{3/2}\kappa^{-1}\ln{\Omega^2}$ as the inflaton, one can obtain the Lagrangian in the Einstein frame
\begin{align}\label{2}
	\mathcal{L}=\sqrt{-\hat{g}}\left[\frac{1}{2\kappa^2}\hat R-\frac{1}{2}(\hat\nabla\phi)^2-\frac{1}{2}e^{-\sqrt{\frac{2}{3}}\kappa\phi}(\hat\nabla\chi)^2-V(\phi,\chi)\right]\;,
\end{align}
where
\begin{align}\label{3}
	V(\phi,\chi)=e^{-2\sqrt{\frac{2}{3}}\kappa\phi}\left[\frac{3\mu^2}{4\kappa^2}\left(e^{\sqrt{\frac{2}{3}}\kappa\phi}-1+\xi\kappa^2\chi^2\right)^2\right]\;.
\end{align}
In a flat Friedmann-Robertson-Walker (FRW) background,  we obtain the following field equations from the Lagrangian~(\ref{2})
\begin{align}\label{4}
\ddot{\phi}+3H\dot{\phi}+\frac{\partial V}{\partial\phi}-\frac{1}{a^{2}}\bigtriangledown^{2}\phi+\frac{\kappa }{\sqrt{6}}e^{-\sqrt{\frac{2}{3}}\kappa\phi}\left(\dot{\chi}^{2}-\frac{1}{a^{2}}\partial^{k}\chi\partial_{k}\chi\right)=0\ ,
\end{align}
\begin{align}\label{5}
	\ddot{\chi}+3H\dot{\chi}+e^{\sqrt{\frac{2}{3}}\kappa\phi}\frac{\partial V}{\partial\chi}-\frac{1}{a^{2}}\bigtriangledown^{2}\chi-\sqrt{\frac{2}{3}}\kappa\left(\dot{\phi}\dot{\chi}-\frac{1}{a^{2}}\partial^{k}\phi\partial_{k}\chi\right)=0\ ,
\end{align}
and
\begin{align}\label{6}
\dfrac{1}{\kappa^{2}} \bigg(\hat{R}_{\mu\nu}-\frac{1}{2}\hat{g}_{\mu\nu}\hat{R}\bigg)=  &\partial_{\mu}\phi\partial_{\nu}\phi+e^{-\sqrt{\frac{2}{3}}\kappa\phi}\partial_{\mu}\chi\partial_{\nu}\chi\nonumber \\
&-\hat{g}_{\mu \nu}\left(\frac{1}{2} \partial^{\alpha}\phi\partial_{\alpha}\phi+\frac{1}{2}e^{-\sqrt{\frac{2}{3}}\kappa\phi} \partial^{\alpha}\chi\partial_{\alpha}\chi+V(\phi,\chi)\right)\ .
\end{align}
Here a dot denotes a derivative with respective to the cosmic time $t$, $a$ is the scale factor, and $H$ is the Hubble parameter. During the inflation which is driven by the inflaton field $\phi$, the $\chi$ field has a negligible effect and thus can be neglected. After inflation, the field $\phi$ behaves as an inflaton with the quadratic potential and oscillates coherently around $\phi=0$. This oscillations lead to that the $\chi$ field may have a tachyonic mass since the square of its effective mass, which has the form
  \begin{align}\label{7}
  	m^2_{\chi,\text{eff}}=\frac{d^2V}{d\chi^2}= e^{-2\sqrt{\frac{2}{3}}\kappa\phi}\left[3\mu^2\xi\left(e^{\sqrt{\frac{2}{3}}\kappa\phi}-1+3\xi\kappa^2\chi^2\right)\right]\;,
  \end{align}
  can be negative. The tachyonic instability causes the parametric resonance of the $\chi$ particles, which excites the production of copious $\chi$ particles of small-momentum modes. Since there are rescatterings between the  $\chi$ particles and the inflaton condensate, the produced $\chi$ particles can knock inflaton quanta out of the condensate into low-momentum modes. So, the rapid growth of the inflaton and $\chi$ fluctuations can be expected, which means that they can become significant GW sources.

The best way to obtain the production of GW during preheating in the Starobinsky model is to perform numerical lattice simulations.  Here, we use the publicly available  package HLattice~\cite{2011Huang} to do simulations. In original  HLattice   the symplectic integrator is used to handle scalar fields with canonical kinetic terms.  Since the scalar fields   considered in this paper have  non-canonical kinetic terms, we modify the HLattice by  adopting the fourth-order Runge-Kutta integrator instead of the symplectic   one. The results are simulated with $N^3=(128)^3$ points.   The lattice length of side $L$ is chosen to   to satisfy $LH_i<2\pi$, which means that all field modes are always within the horizon, where $H_i$ ($\simeq 3.87\times10^{-6}M_{p}$) is the Hubble parameter when the preheating begins.  Different values of $LH_i$ are chosen for the different values of $\xi$. For example, when $\xi=5$ we set $LH_i$ to be $5$, which indicates that the minimum of the comoving wave number $k$ is $k_{\mathrm{min}}=1.25H_i$.  Since all field modes are sub-horizon, we will initialize the fluctuations of $\phi$ and $\chi$ fields to be those in the Bunch-Davies vacuum.
In addition, we set a UV cutoff $k_{\mathrm{UV}}$ on $k$ when initializing the fluctuations, i.e.,  $k_{\mathrm{UV}}= 30 k_{\rm min}$ when $\xi=5$, which is larger than the maximum resonance frequency in order to prevent it from affecting the results. Furthermore, we have checked that the results are independent of the values of  $k_{\mathrm{UV}}$. In the original HLattice, the produced results are also limited to be in the regions $k\in [k_{\mathrm{min}}, k_{\mathrm{UV}}]$. Here,  we relax this limitation to include  some momenta larger than $k_{\mathrm{UV}}$. 
This operation is similar to  what was done in \cite{Figueroa20172}. The initial conditions of the homogeneous part of $\phi$ field during preheating are determined by the time when the inflation ends, while for the $\chi$ field the homogeneous part is initialized to zero. We stop the simulation when the density spectrum of the inflaton quanta and the $\chi$ field do not change appreciably.

\section{Production of gravitational waves}
\label{sec3}
Now we study the production of GWs during preheating in the Starobinsky inflationary model. GW can be represented by the transverse traceless part of the spatial metric disturbance of the FRW background
  \begin{align}\label{8}
	ds^2=g_{\mu\nu}dx^\mu dx^\nu
	=-dt^2+a(t)^2(\delta_{ij}+h_{ij})dx^idx^j
\end{align}
with $\partial_{i}h_{ij}=h_{ii}=0$. The perturbation $h_{ij}$ corresponds to two independent tensor degrees of freedom and satisfies the equation of motion
\begin{align}\label{9}
\ddot h_{ij}+3H\dot h_{ij}-\frac{1}{a^2}\nabla^2 h_{ij}=\frac{2\kappa^2}{a^{2}}\Pi_{ij}^{TT} \ ,
\end{align}
where the source term $\Pi_{ij}^{TT}$ is the transverse-traceless part of the anisotropic stress $\Pi_{ij}$, which is defined as
\begin{align}\label{10}
\Pi_{ij}\equiv T_{ij}-\left \langle p \right\rangle g_{ij}
\end{align}
with $T_{ij}$ and $p$ being the energy-momentum tensor and  the pressure of  the system, respectively.
For the model considered in this paper the energy-momentum tensor has the form
\begin{align}\label{11}
	T_{\mu \nu} = &\partial_{\mu}\phi\partial_{\nu}\phi+e^{-\sqrt{\frac{2}{3}}\kappa\phi}\partial_{\mu}\chi\partial_{\nu}\chi\nonumber \\
	&-g_{\mu \nu}\left(\frac{1}{2} \partial^{\alpha}\phi\partial_{\alpha}\phi+\frac{1}{2}e^{-\sqrt{\frac{2}{3}}\kappa\phi} \partial^{\alpha}\chi\partial_{\alpha}\chi+V(\phi,\chi)\right) \ ,
\end{align}
and the anisotropic stress can be expressed as
\begin{align}\label{12}
	\Pi_{ij}=\partial_{i}\phi\partial_{j}\phi+e^{-\sqrt{\frac{2}{3}}\kappa\phi}\partial_{i}\chi\partial_{j}\chi-\frac{1}{3}\delta_{ij}\left( \partial^{k}\phi\partial_{k}\phi+e^{-\sqrt{\frac{2}{3}}\kappa\phi} \partial^{k}\chi\partial_{k}\chi\right) \ .
\end{align}
From the above expression, one can obtain  the transverse-traceless part of the anisotropic stress
\begin{align}\label{13}
	\Pi_{ij}^{TT}=\left[\partial_{i}\phi\partial_{j}\phi+e^{-\sqrt{\frac{2}{3}}\kappa\phi}\partial_{i}\chi\partial_{j}\chi\right]^{TT} \ .
\end{align}

In our numerical calculation, it is more convenient to work in the Fourier space. Using the convention
\begin{align}\label{14}
f(\mathbf{k}) = \frac{1}{(2\pi)^{3/2}} \int d^{3}\mathbf{x} e^{i\mathbf{k}\cdot \mathbf{x}}f(\mathbf{x}) \ ,
\end{align}
one can obtain that the GW equation~(\ref{9}) in Fourier space has the form
\begin{align}\label{15}
\ddot h_{ij}(t,\mathbf{k})+3H\dot h_{ij}(t,\mathbf{k})-\frac{k^{2}}{a^2} h_{ij}(t,\mathbf{k})=\frac{2\kappa^2}{a^{2}}\Pi_{ij}^{TT}(t,\mathbf{k})
\end{align}
where $k=|\mathbf{k}|$. After introducing the transverse-traceless projection operator~\cite{Bellido2008}:   $\Lambda_{ij,lm}(\mathbf{k})=P_{il}(\mathbf{k})P_{jm}(\mathbf{k})-\frac{1}{2}P_{ij}(\mathbf{k})P_{lm}(\mathbf{k})$, where $
P_{ij}(\mathbf{k})=\delta_{ij}-\frac{k_i k_j}{k^2}$ is the spatial projection
operators,  we achieve the transverse-traceless part of $\Pi_{ij}$ in the momentum space
\begin{align}\label{16}
\Pi^{TT}_{ij}(\mathbf{k})=\Lambda_{ij,lm}(\mathbf{k})\Pi_{lm}(\mathbf{k})
\end{align}
It is easy to see that  $k_{i}\Pi_{ij}^{TT}(t,\mathbf{k})=\Pi_{ii}^{TT}(t,\mathbf{k})=0$ in any time $t$.
Using this projection operator,  one can define a new tensor $u_{ij}$ which satisfies the following relation in the momentum space
\begin{align}\label{17}
h_{ij}(t,\mathbf{k})=\Lambda_{ij,lm}(\mathbf{k})u_{lm}(t,\mathbf{k}) \ .
\end{align}
 Then the GW equation~(\ref{9}) can be re-written as
\begin{align}\label{18}
\ddot u_{ij}+3\frac{\dot a}{a}\dot u_{ij}-\frac{1}{a^2}\nabla^2 u_{ij}=\frac{2\kappa^2}{a^2}[\partial_i\phi\partial_j\phi+e^{-\sqrt{\frac{2}{3}}\kappa\phi}\partial_i\chi\partial_j\chi] \ .
\end{align}
 We assume no GWs at the beginning of preheating, so $u_{ij}$ and its derivative are  initialized to be zero.

The GW energy density is given by
\begin{align}\label{19}
\rho_{gw}=\sum\limits_{i,j}\frac{1}{4 \kappa^{2}}\langle \dot h^2_{ij}\rangle \ .
\end{align}
Here $\langle\cdot\cdot\cdot\rangle$ is the spatial average.
According to the Parseval's theorem,
the GW energy density can be expressed as an integral in momentum space
\begin{align}\label{20}
\rho_{\rm gw}&=\frac{1}{L^{3}} \frac{1}{4 \kappa^{2}}\int d^{3}\mathbf{k}  \dot{h}_{ij}(t,\mathbf{k})\dot{h}_{ij}^{\ast}(t,\mathbf{k})\nonumber \\
&=\frac{1}{L^{3}} \frac{1}{4 \kappa^{2}}\int d^{3}\mathbf{k}  \Lambda_{ij,lm}(\mathbf{k})\dot {u}_{lm}(t,\mathbf{k})\Lambda_{ij,rs}\dot {u}^{\ast}_{rs}(t,\mathbf{k})\nonumber \\
&=\frac{1}{L^{3}} \frac{1}{4 \kappa^{2}}\int d^{3}\mathbf{k}  \Lambda_{lm,rs}(\mathbf{k})\dot {u}_{lm}(t,\mathbf{k})\dot {u}^{\ast}_{rs}(t,\mathbf{k})\ ,
\end{align}
where $L$ is the length of one side of the lattice.
The corresponding GW spectra, normalized to the critical energy density $\rho_{c}$,  can be obtained through
\begin{align}\label{21}
\Omega_{\rm gw}(k)\equiv \frac{1}{\rho_{c}}\frac{ d \rho_{\rm gw}}{d\ln{k}}=\frac{\pi k^3}{3H^2L^3} \int\frac{d\Omega_k}{4\pi}\Lambda_{ij,lm}(\mathbf{k})\dot u_{ij}(\mathbf{k})\dot u^{\ast}_{lm}(\mathbf{k})\ .
\end{align}

\begin{figure}
	\centering
	\includegraphics[width=0.49\textwidth ]{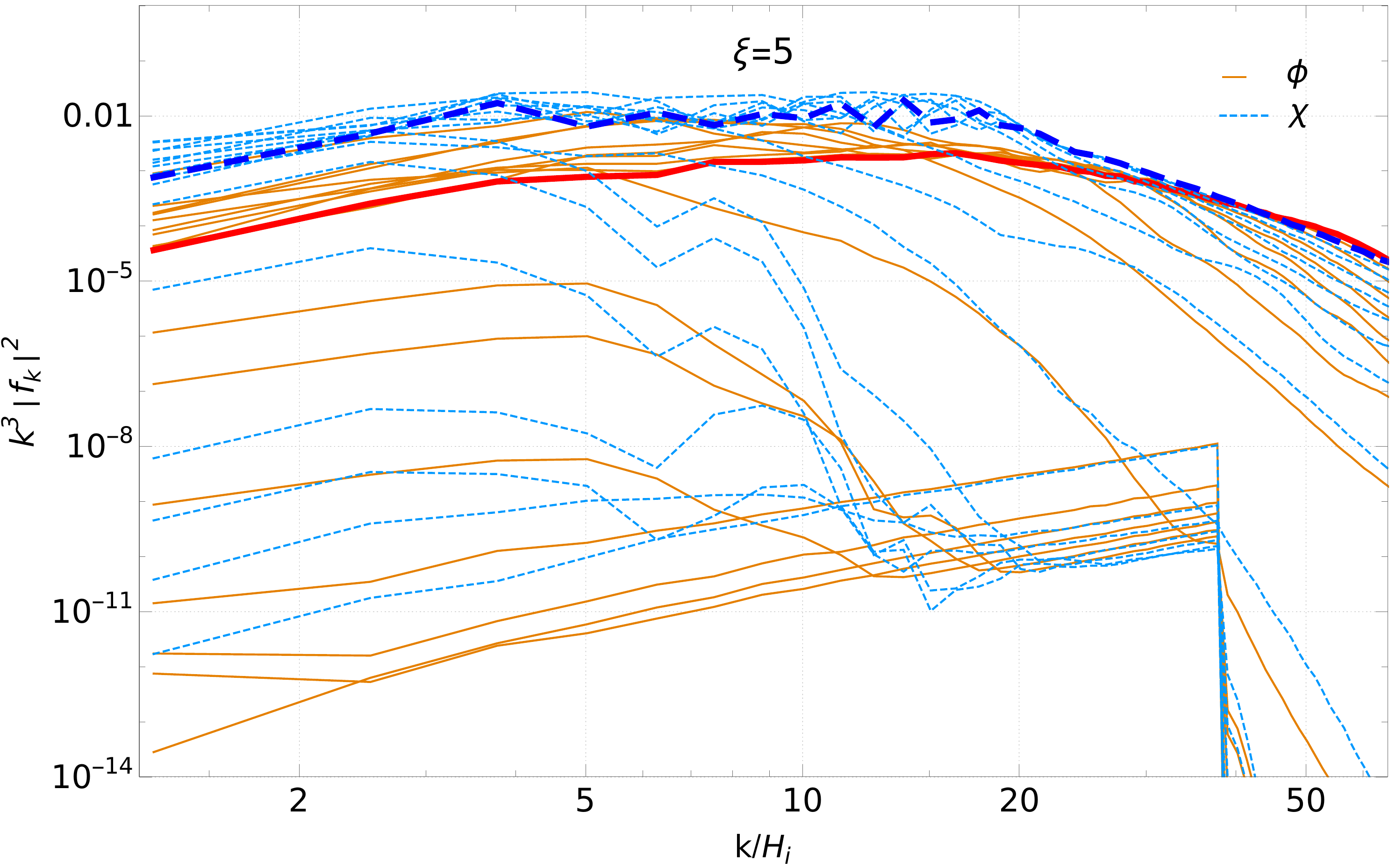}
	\includegraphics[width=0.49\textwidth ]{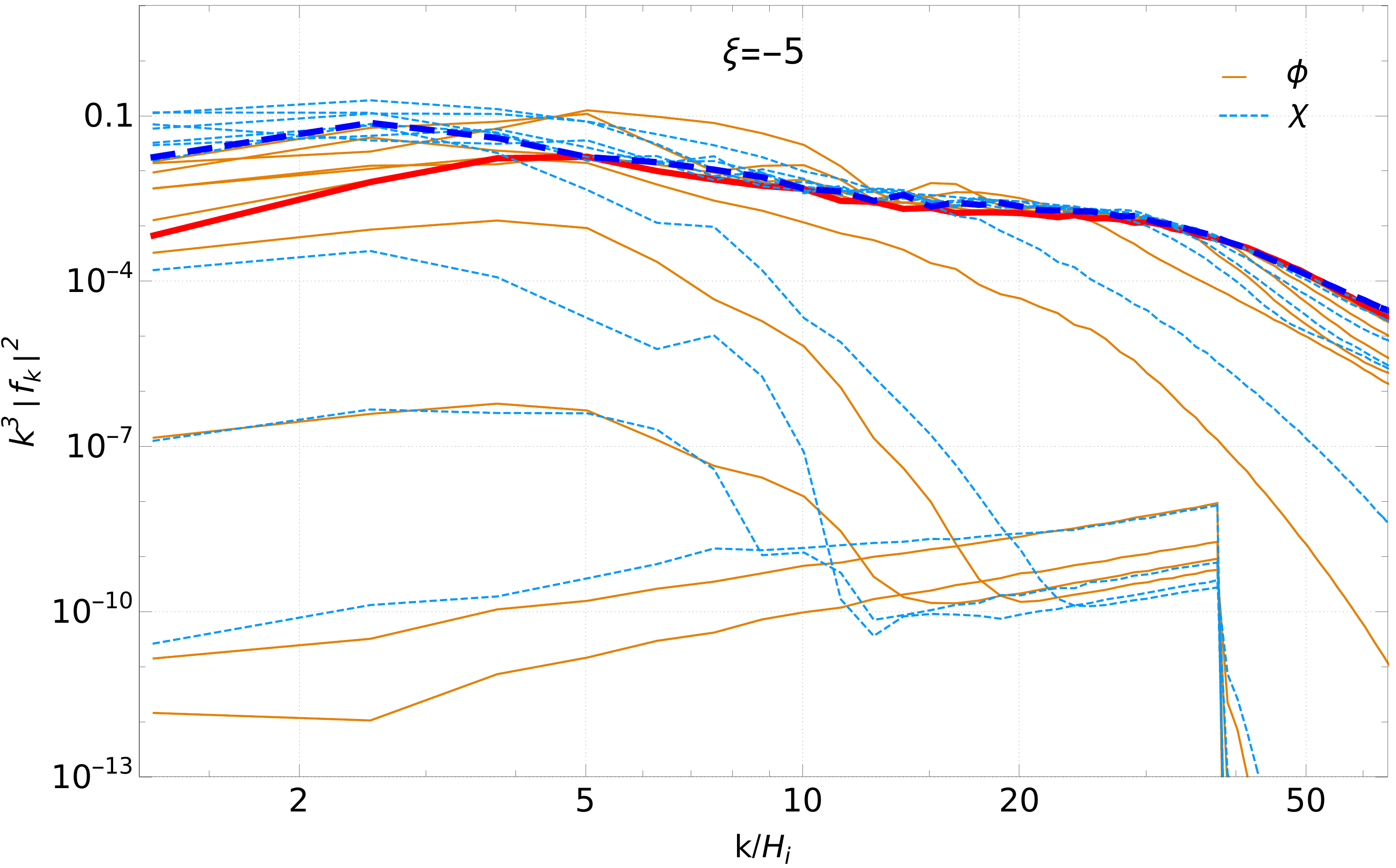}
\caption{\label{fig1} The evolutions of the spectra of scalar fields as a function of  $k/H_{i}$ with time $t$.  The spectra from bottom to up are plotted with the time interval  $\Delta t=0.1875 H^{-1}_{i}$.  Bold lines  show the final results. }
\end{figure}

 Since the produced GWs during preheating are determined by the spectra of scalar fields, we first investigate the evolutions of $\chi$ and $\phi$ and  show their spectra in Fig.~\ref{fig1} where only $\xi=-5$ and $5$ are considered as examples. It is easy to see that due to the parametric resonance  $\chi$ particles are produced rapidly.  When $\xi=-5$ the production is quicker than when  $\xi=5$, which results from that  all modes of the $\chi$ field with $k^2/a^2<|m^2_{\chi,\text{eff}}|$  are already tachyonic when the preheating begins~\cite{2019Fu}. After the increase of the $\chi$ spectrum, the spectrum of the $\phi$ field begins to increase since inflaton particles are knocked out of the inflaton condensate through the  rescattering between  $\chi$ particles and the inflaton condensate.  Finally,   both the inflaton field and the auxiliary  field  become the effective GW sources.

\begin{figure}
	\centering
	\includegraphics[width=0.49\textwidth ]{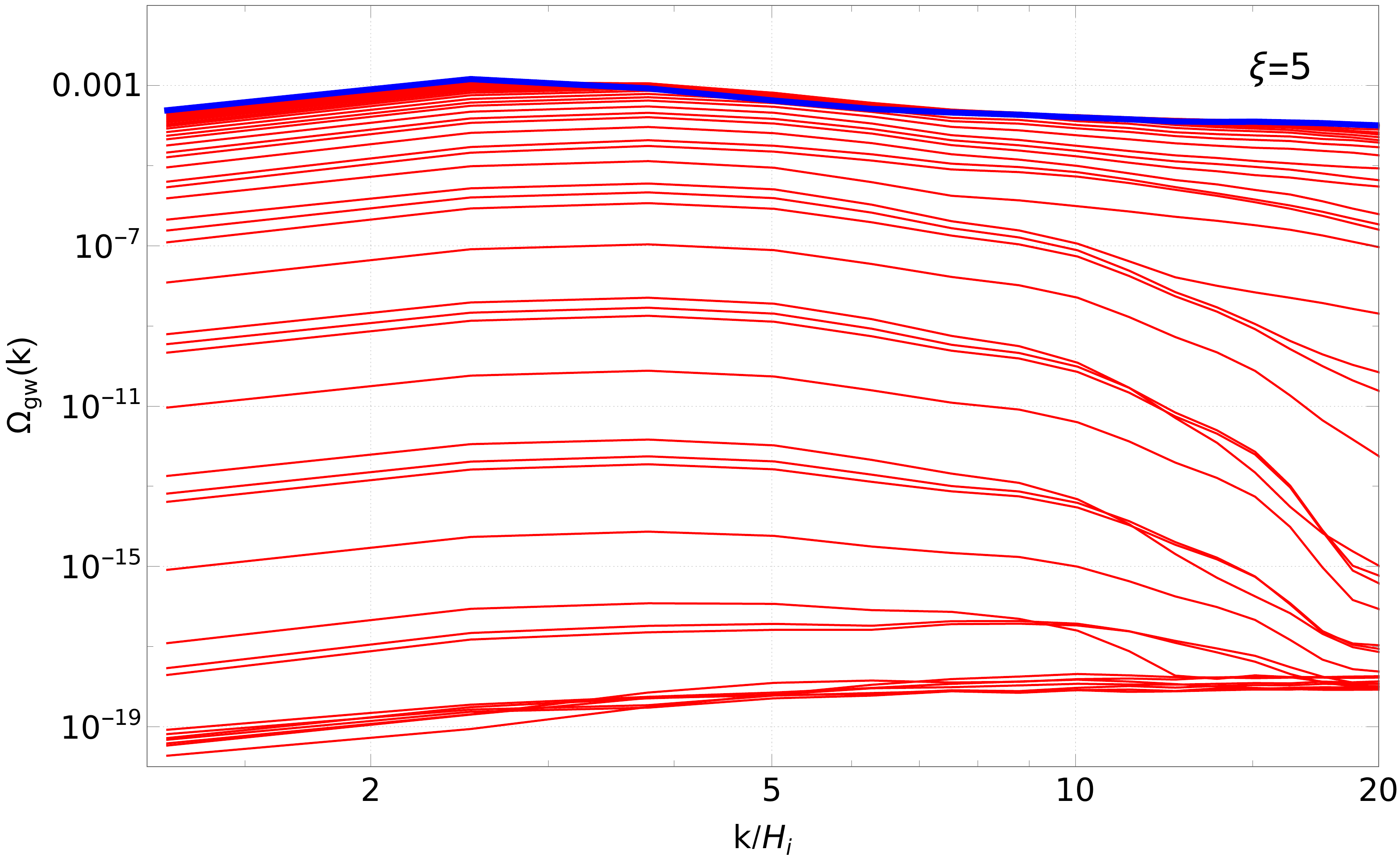}
	\includegraphics[width=0.49\textwidth ]{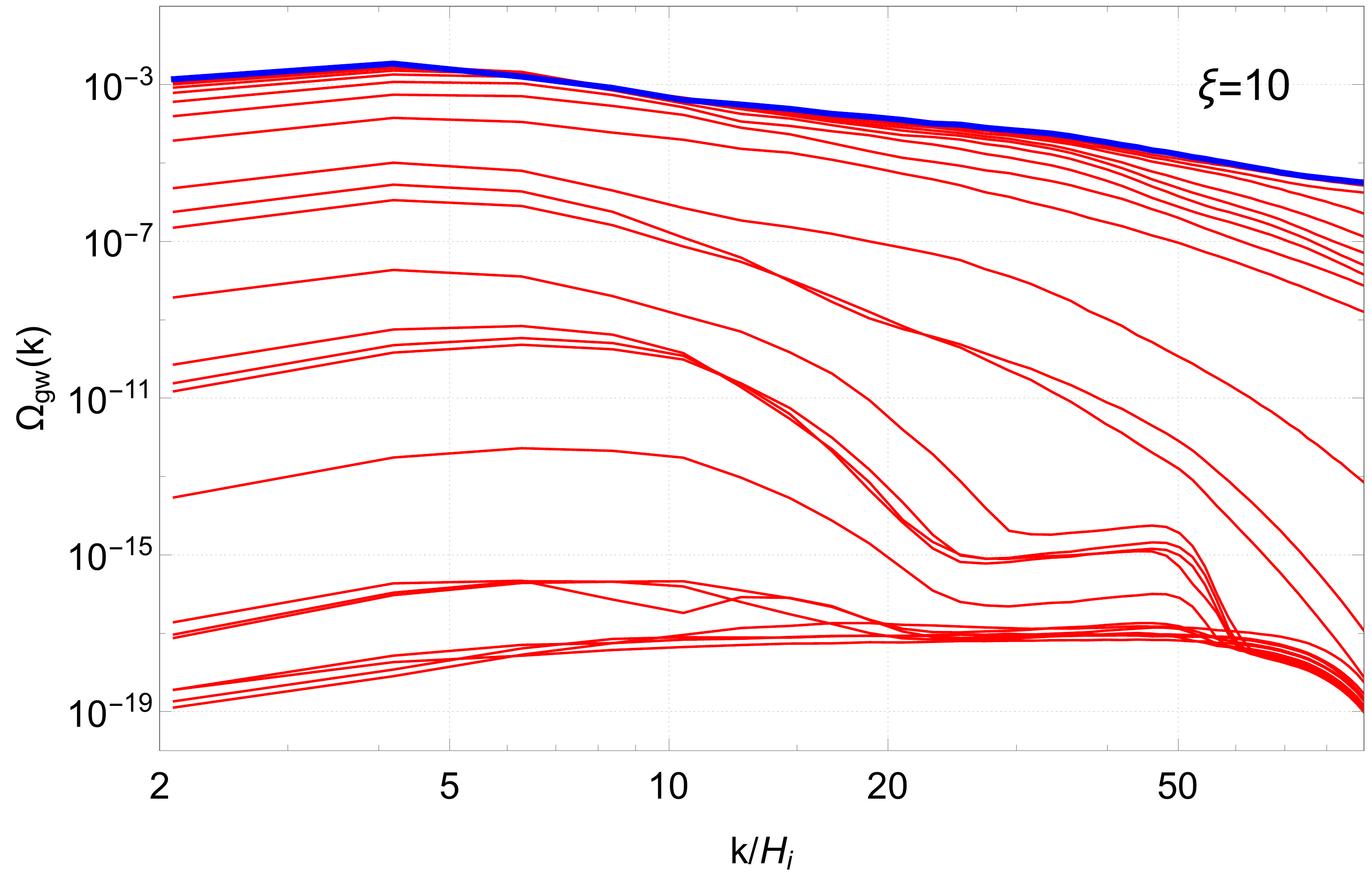}
	\includegraphics[width=0.49\textwidth ]{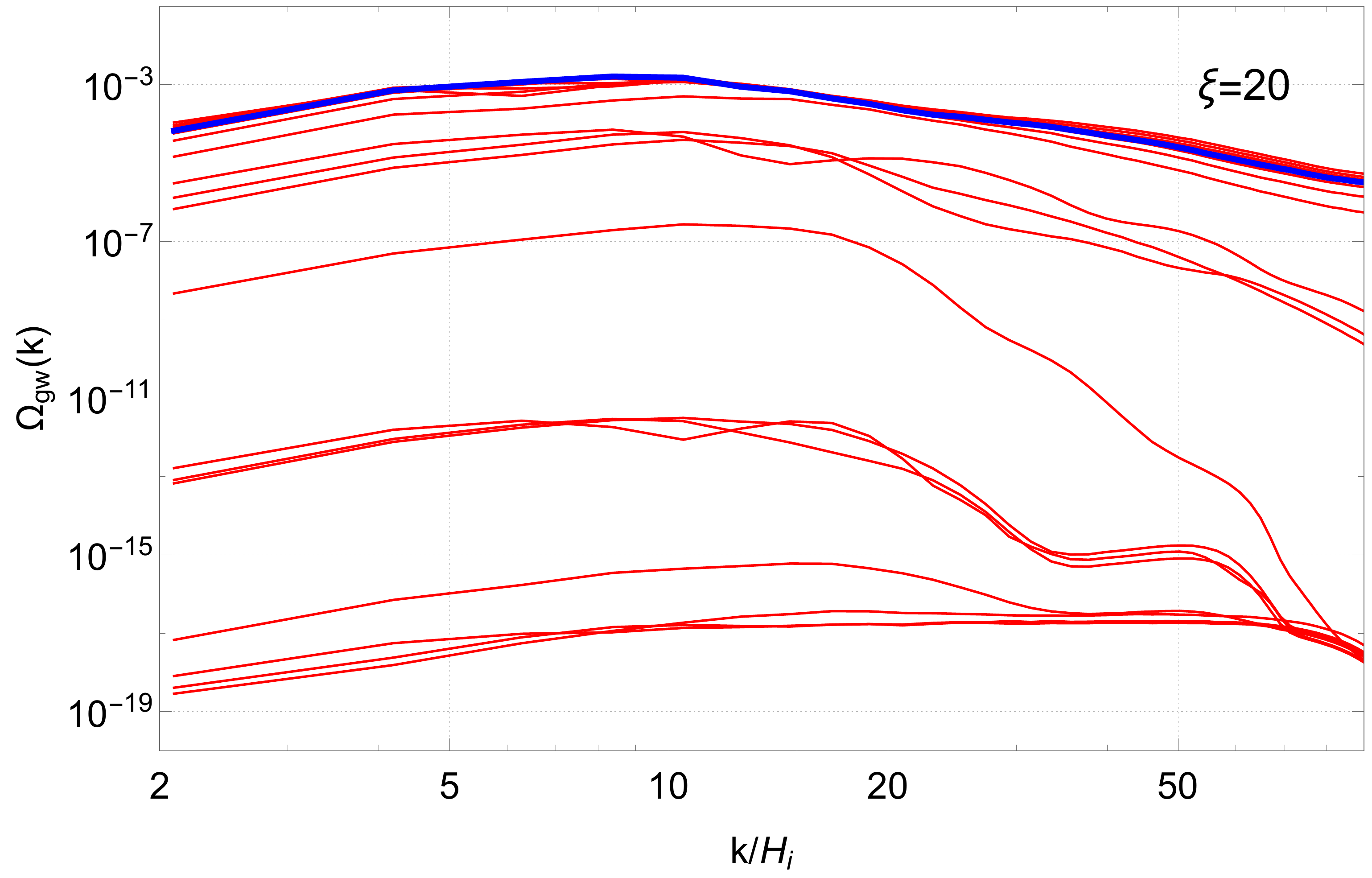}
	\includegraphics[width=0.49\textwidth ]{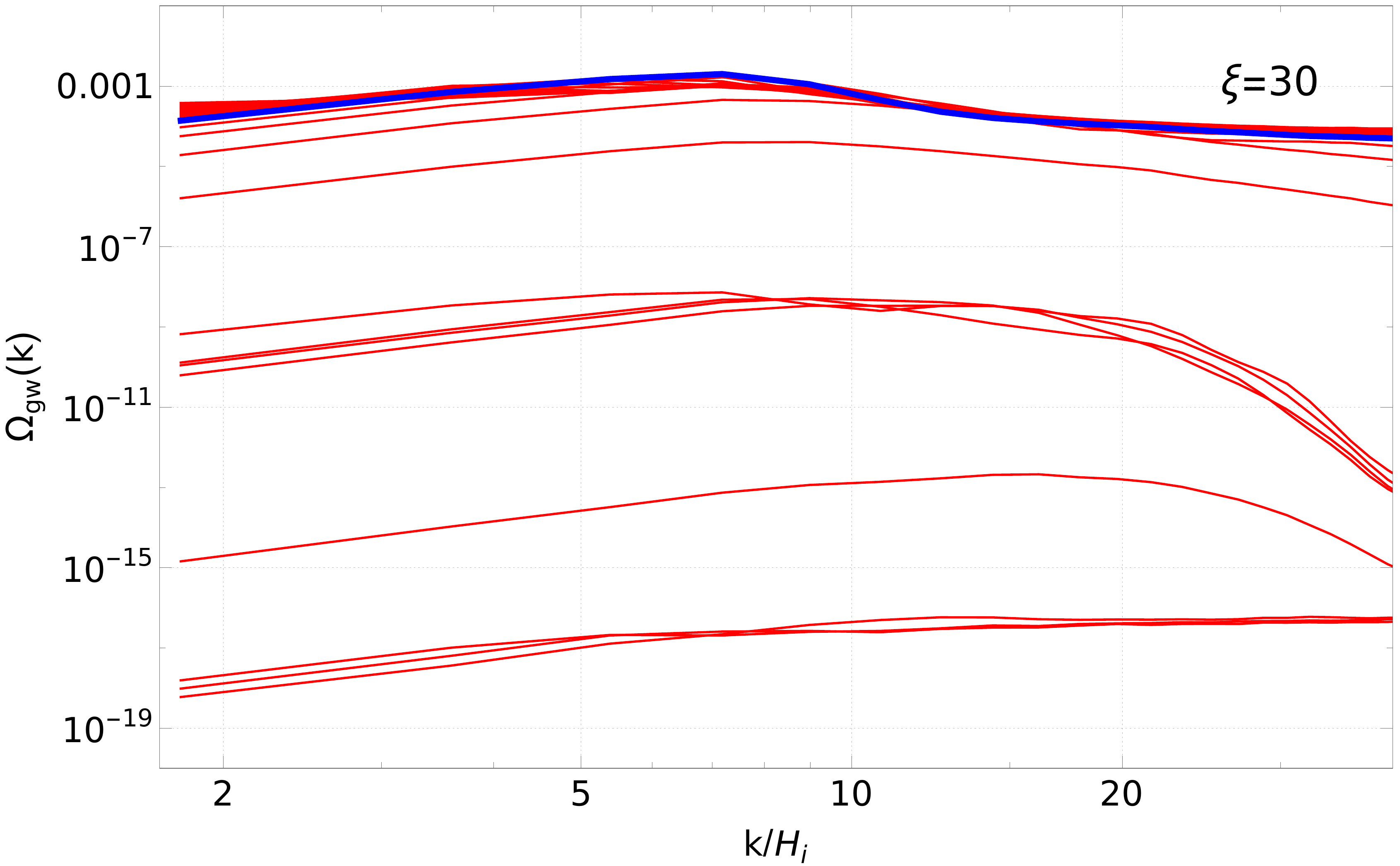}
\caption{\label{fig2} The evolutions of the density spectra of GWs as a function of  $k/H_{i}$ with time $t$ for $\xi>0$.  The spectra from bottom to up are plotted with the time interval  $\Delta t=0.0625 H^{-1}_{i}$ , with Blue line corresponding to the final result. }
\end{figure}

Fig.~\ref{fig2} shows the evolutions of the GW density spectra from the lattice simulation with $\xi=5, 10$, $20$ and $30$. The outputs are plotted from bottom to top
per same time step $\Delta t=0.0625 H^{-1}_{i}$, where $H_{i}$ denotes the value of Hubble rate when inflation ends.
 One can see clearly that the low frequency modes increase rapidly  as parametric resonance begins, and then the growth of the higher frequency modes is very fast due to the nonlinear effect. The amplitude increases faster and faster with the increase of $\xi$, which results from the fact that the resonant bands become broader and broader with the increase of $\xi$. The blue curves, which show the final GW spectra, indicate that the maximum value of $\Omega_{\rm gw}(k)$, $\Omega_{\rm gw, max}$, is not a monotonous function of $\xi$, since when $\xi=10$ the value is larger than the ones of $\xi=5$ and $\xi=30$, which is different from the case of non-minimally coupled inflation model in which the GW spectra produced in preheating increase with the increase of the coupling parameter~\cite{Fu2018}.

\begin{figure}
	\centering
	\includegraphics[width=0.49\textwidth ]{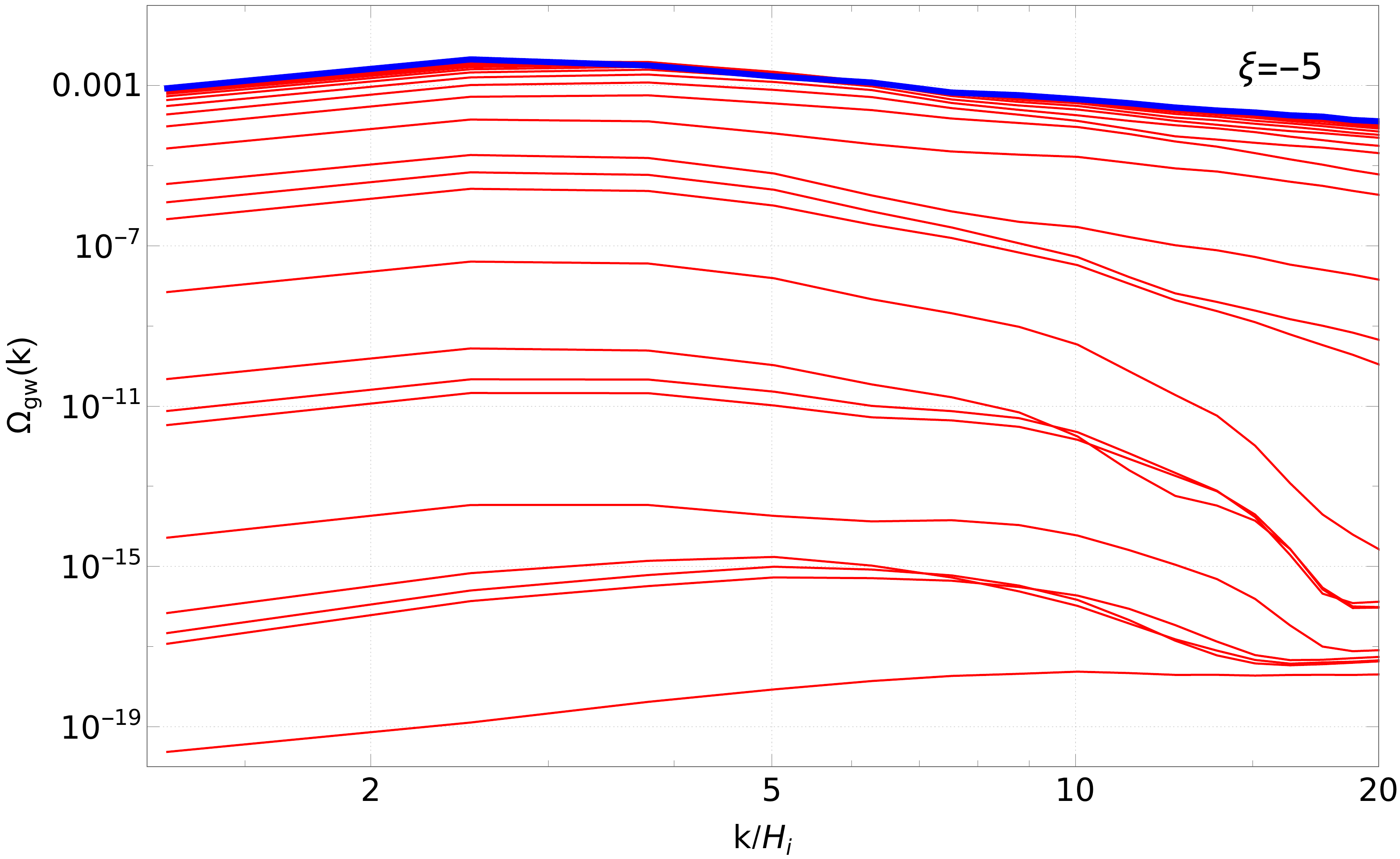}
	\includegraphics[width=0.49\textwidth ]{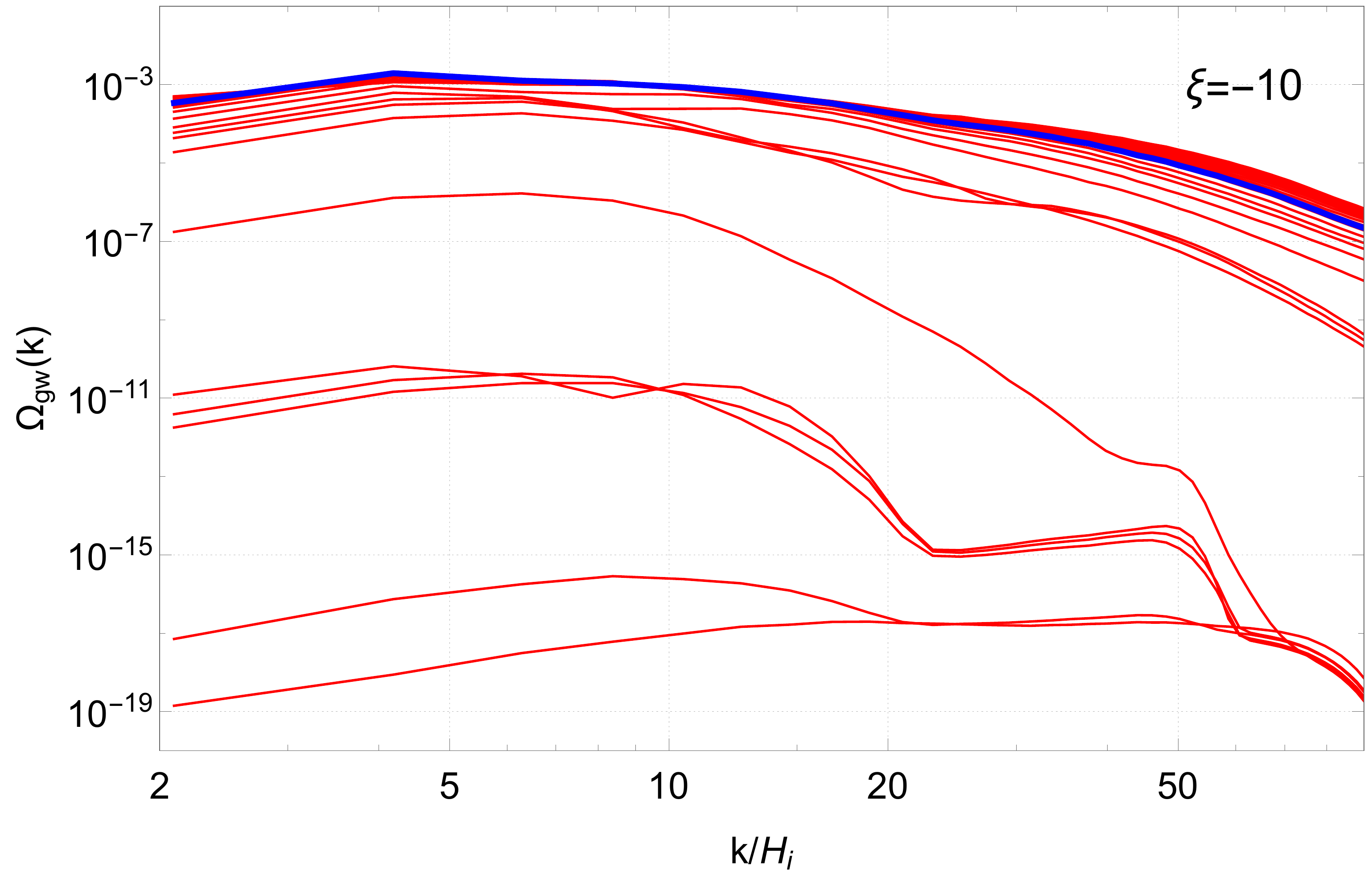}
	\includegraphics[width=0.49\textwidth ]{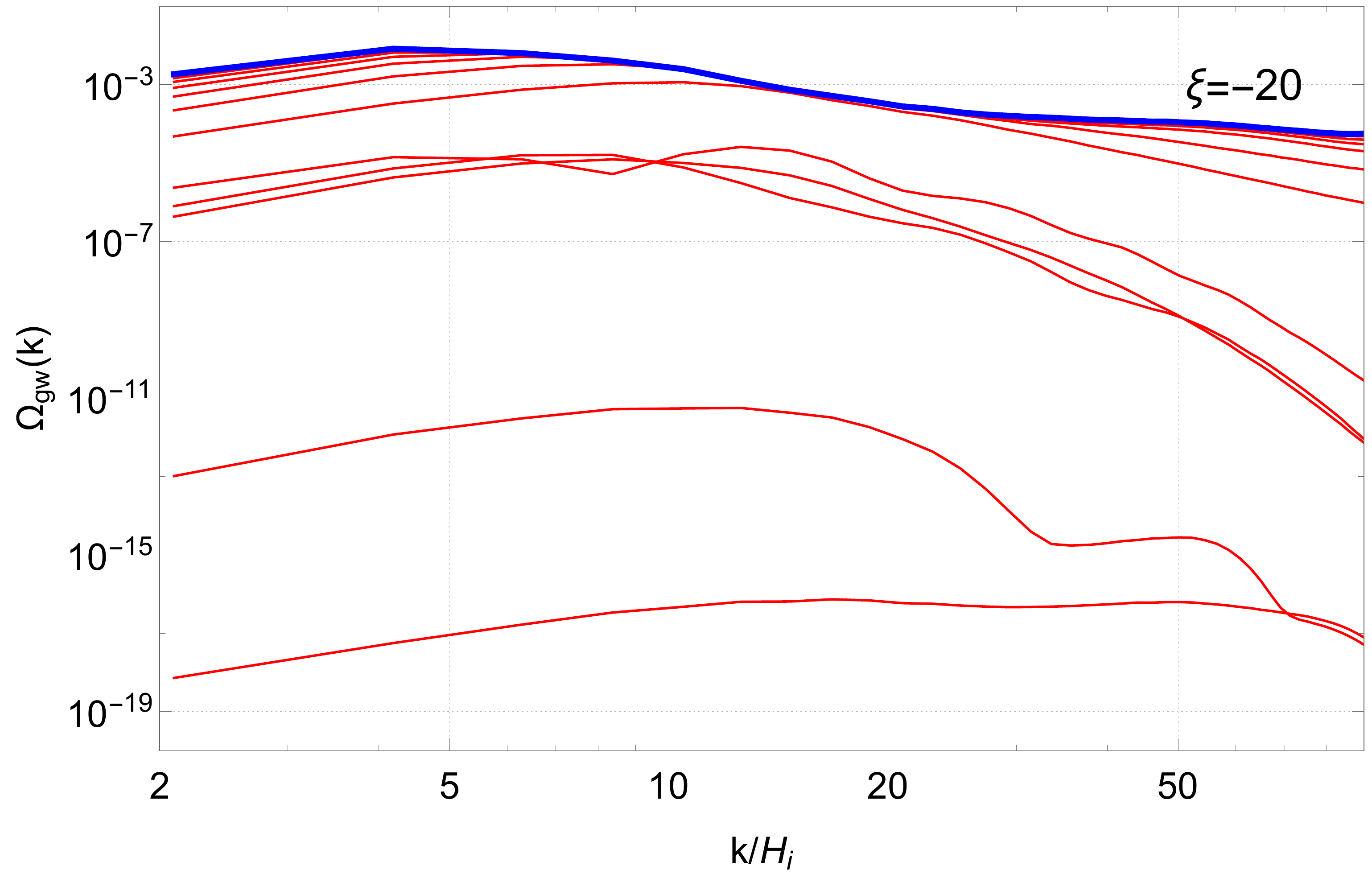}
	\includegraphics[width=0.49\textwidth ]{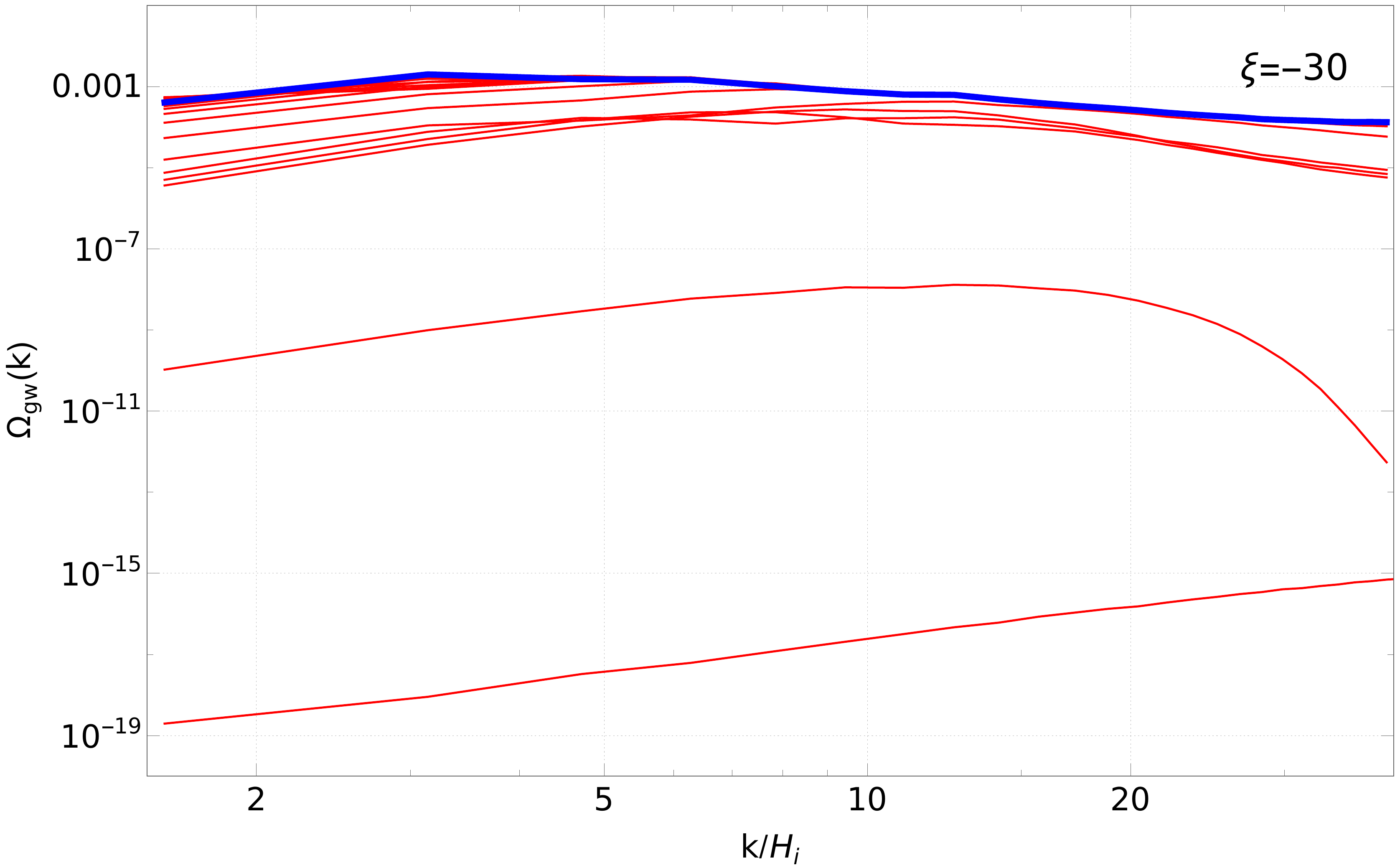}	
	\caption{\label{fig3} The evolutions of the density  spectra of GWs as a function of  $k/H_{i}$ with time $t$ for $\xi<0$. The spectra from bottom  to up are plotted with the time interval  $\Delta t=0.0625 H^{-1}_{i}$ , with Blue line corresponding to the final result. }
\end{figure}

Fig.~\ref{fig3} give the evolutions of the GW density  spectra for the case of negative $\xi$. We take four different values: $\xi=-5$, $-10$, $-20$ and $-30$. The results are very similar with the ones shown in Fig.~\ref{fig2}. This is because the resonant bands become broad with the increase of $\left| \xi\right|$.  But in the case of $\xi<0$, $\Omega_{\rm gw}$ begins to grow once the inflation ends and grows faster than the case of $\xi>0$. The reason is that  when $\xi<0$, all modes of the $\chi$ field with $k^2/a^2<|m^2_{\chi,\text{eff}}|$  are already tachyonic when the preheating begins~\cite{2019Fu}. In addition, we find when $\xi<0$,  $\Omega_{\rm gw, max}$ with $\xi=-20$ is larger than the one from other three cases.

\begin{figure}
	\centering
	\includegraphics[width=0.49\textwidth ]{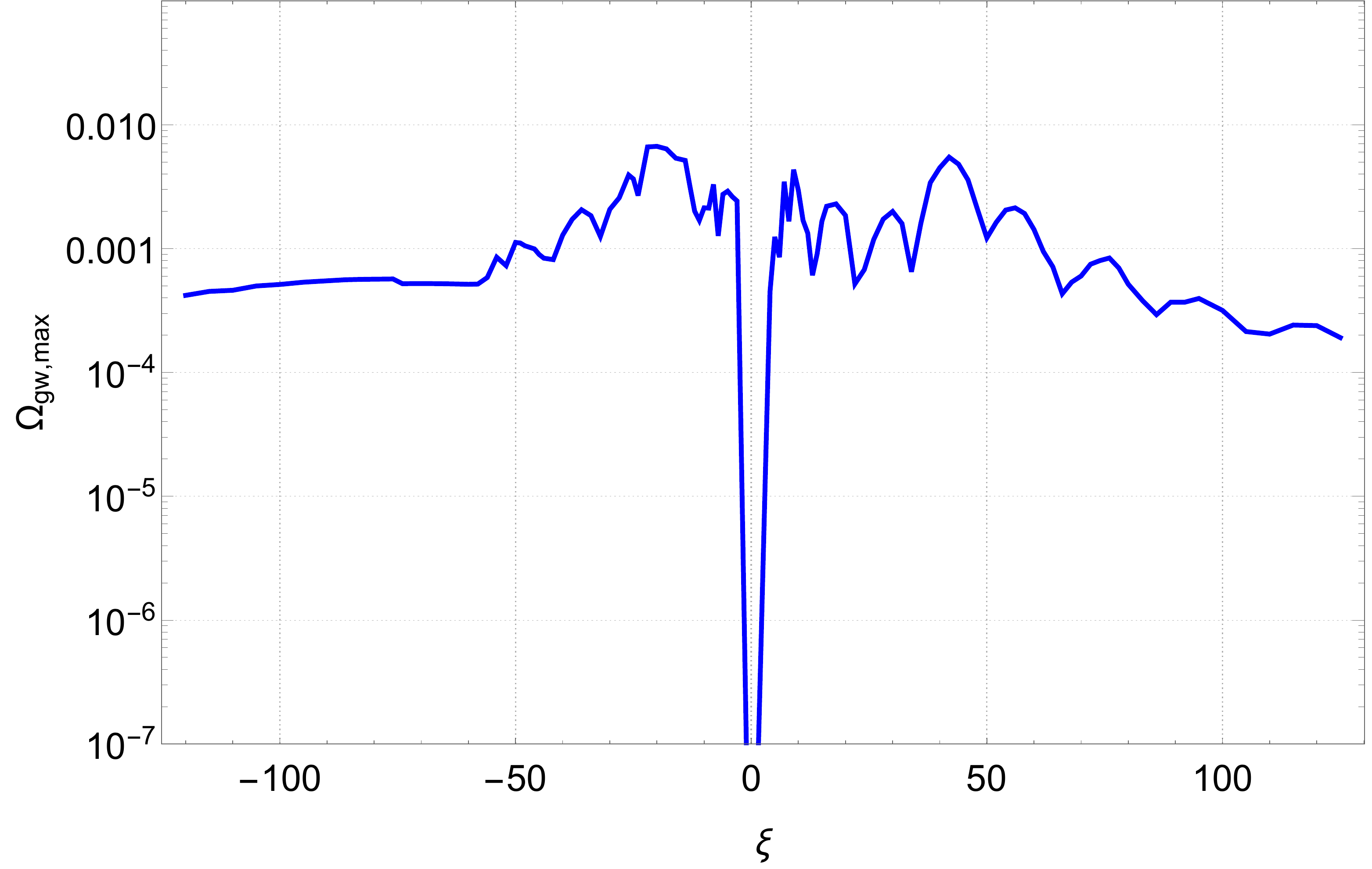}
\caption{\label{fig4}  The amplitude for the final GW density spectrum peak as a function of $\xi$. }\end{figure}

Although Figs.~\ref{fig2} and \ref{fig3} do not indicate  a simple relation  between $\Omega_{\rm gw, max}$ and  the coupling constant $\xi$,  there, however, indeed exists a subtle relation as we will now demonstrate. For this purpose, we plot the evolution of $\Omega_{\rm gw, max}$ as a function of $\xi$ in Fig.~\ref{fig4}. We find that  when $\xi$ is very small the production of  GWs is very weak, which is agreement with the result obtained  in ~\cite{Starobinsky1981} where it was found that the direct generation of GWs after the end of inflation  is suppressed in the Starobinsky model.  One can see that the maximum of $\Omega_{\rm gw}$ is about $6.67\times10^{-3}$, which occurs at $\xi \simeq -20$ . In the case of $\xi>0$, the maximum of $\Omega_{\rm gw}$ is about $5.46\times10^{-3}$ at $\xi \simeq 42$. It is easy to see that $\Omega_{\rm gw,max}$ first increases fastly with the increase of $|\xi|$,  and then reaches the maximum after several oscillations. Finally it decreases with slight oscillation with the further increase of $\left|\xi\right|$.

Since we are interested in the energy density of GWs today, the GW spectra generated in the preheating era need to be transformed to the present-day case.  The present scale factor compared to that at the time when  GW production stops can  be expressed as
\begin{align}\label{22}
	\frac{a_{*}}{a_{0}}=\left(\frac{a_{*}}{a_{j}}\right)^{1-\frac{3}{4}\left(1+w\right)}\left(\frac{\bar{g}_{j}}{ \bar{g}_{0}}\right)^{-1/12}\left(\frac{\rho_{r,0}}{\rho_{*}}\right)^{1/4} \ ,
\end{align}
where subscripts $*$ , $j$ and $0$  denote  the time when GW production stops, thermal equilibrium is established and today, respectively.
Here, $w$ is the mean equation of state of the cosmic energy from $t_{*}$ to $t_{j}$, $\rho_{r}$  is the radiation energy density, $\bar{g}$ is the number of effectively relativistic degrees of freedom, and $\rho_*$ is the total energy density of scalar fields.
So the present-day GW physical frequency is given by
\begin{align}\label{23}
	f=\frac{k}{2 \pi a_{0}}=\frac{k}{ a_{*}\rho^{1/4}_{*}}\left(\frac{a_{*}}{a_{j}}\right)^{1-\frac{3}{4}\left(1+w\right)}\times  (4.4 \times 10^{10}) ~Hz  \ .
\end{align}
Here, we take the reheating temperature to be about $3.1 \times 10^{9} \rm Gev$~\cite{Gorbunov:2010bn},  which means that one can take $\bar{g}_{j}/\bar{g}_{0}=106.75/3.36 \simeq 31$ in  analysis. The transformed function to obtain today's GW amplitude is~\cite{Easther2}
\begin{equation}\label{24}
	\Omega_{\rm gw,0}(f)h^{2}=\Omega_{\rm gw}(f)\left(\frac{\bar{g}_{j}}{\bar{g}_{0}}\right)^{-1/3}\Omega_{ r,0}h^{2}\left(\frac{a_{*}}{a_{j}}\right)^{1-3w } \ ,
\end{equation}
where $\Omega_{ r,0}h^{2}=h^{2}\rho_{ r,0}/\rho_{c,0}=4.15\times 10^{-5}$ is the abundance of radiation today and $h$ is the present dimensionless Hubble constant. Thus, the present-day GW  energy density  has the form
\begin{align}\label{25}
	\Omega_{\rm gw,0}h^{2}=\int d \ln f \ \Omega_{\rm gw,0}(f)h^{2}\ .
\end{align}
  During reheating the mean equation of state $w$ varies with time, but its form is unknown. In our analysis  we assume $w=1/3$ between $t_*$ and $t_j$ for simplicity. This assumption is reasonable  since in~\cite{2019Fu} it has been found that after preheating  the value of $w$ is close to $0.3$. 

 Since the GW energy density scales like radiation, it may be constrained by CMB and BBN measurements of the total radiation density in species beyond the standard model~\cite{2018Peter}. If we assume that beyond the standard model all extra radiation density today during the formation of the CMB is comprised of GWs, the total GW energy density  is constrained from a bound on $\Delta N_{\rm eff}$~\cite{2000Michele}
\begin{align}\label{26}
	\frac{\Omega_{\rm gw,0}h^{2}}{\Omega_{\gamma,0}h^{2}}=\frac{7}{8}\left(\frac{4}{11}\right)^{4/3}\Delta N_{\rm eff} \ ,
\end{align}
where $\Omega_{\gamma,0}h^{2}=2.47 \times 10^{-5}$ is the present energy density of photons, $N_{\rm eff}$ is the effective extra relativistic degrees of freedom and $\Delta N_{\rm eff}= N_{\rm eff}-3.046$. The Planck 2018 results~\cite{2018Plan} limit $\left|\Delta N_{\rm eff} \right| \lesssim 0.33$, which requires that  the GW energy density must satisfy $\Omega_{\rm gw,0}h^{2} \lesssim 1.85 \times 10^{-6}$. The next-generation CMB experiments, such as CMB-S4, will probe $\Delta N_\mathrm{eff} \leq 0.03$ at $1\sigma$ CL and $\Delta N_\mathrm{eff} \leq 0.06$ at $2\sigma$ CL~\cite{Abazajian:2019eic}, which potentially constrains the energy density to be
\begin{align}\label{27}
\Omega_{\rm gw,0} h^2 \lesssim 1.68 - 3.36 \times 10^{-7}
\end{align}

\begin{figure}
	\centering
	\includegraphics[width=0.5\textwidth ]{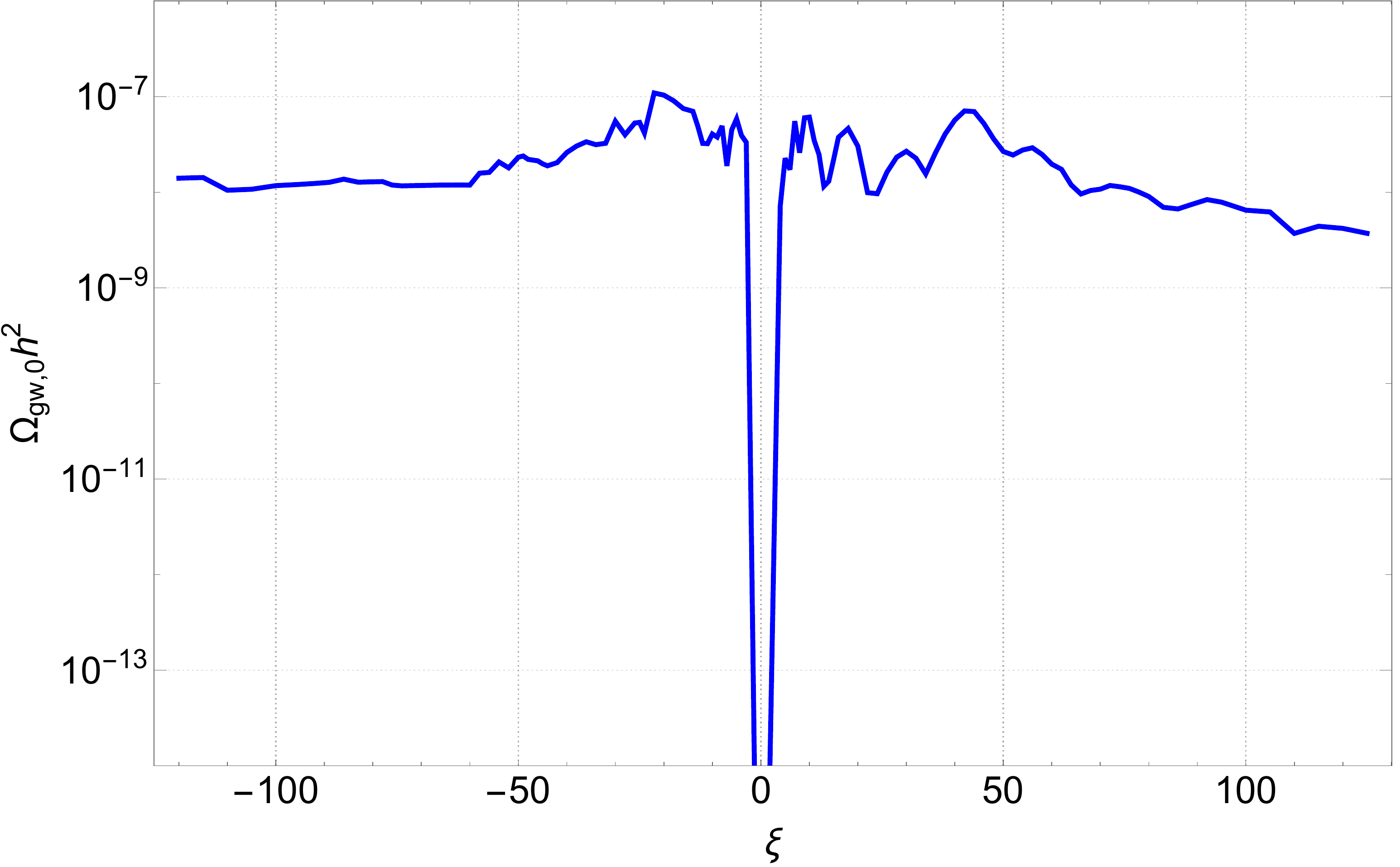}
	\caption{\label{fig5} The total GW energy density today as a function of $\xi$ .}
\end{figure}

 In Fig.~\ref{fig5}, we show the total GW energy density today as a function of $\xi$.   One can see that the maximum of energy density is $1.09\times10^{-7}$, which occurs at $\xi \simeq -21$. Thus,  the GWs produced during  preheating  in the Starobinsky inflationary model satisfy the Planck limit and  the next-generation CMB experiment constraints.

 \section{Conclusion}

The Starobinsky inflationary model satisfies the  latest CMB observations very well. In this inflationary model, to achieve the preheating process requires an auxiliary scalar field which couples non-minimally with the gravity.  This  auxiliary scalar field becomes an effective GW source due to the parametric resonance. Furthermore, the rescattering between auxiliary scalar field and inflation field can knock copious inflaton particles out of the inflaton condensate, which results in that the inflaton field also becomes an effective GW source.  In this paper, using the lattice simulation, we study the production of GWs during preheating in the Starobinsky model. We find that the GW spectrum $\Omega_{\rm gw}$  grows fast with the increase of the absolute value of the coupling parameter $\xi$. This is because the resonant bands become broad with the increase of $|\xi|$.  When $\xi<0$, $\Omega_{\rm gw}$ begins to grow once the inflation ends and grows faster than the case of $\xi>0$.  $\Omega_{\rm gw}$  reaches the maximum at  $\xi\simeq -20$ ($\xi\simeq 42$ for the case $\xi>0$) and then decreases with slight oscillation. By assuming that  all extra radiation density during the CMB formation is comprised of GWs, we investigate the constraint on the total GW energy density from observations  and  find that  the GWs  producing during preheating in the Starobinsky mode satisfy the limits from the Planck and next-generation CMB experiments.

\begin{acknowledgments}
We appreciate very much the insightful comments and helpful suggestions by anonymous referees.
This work was supported in part by the NSFC under Grants No. 11435006, No. 11690034, No. 11805063, and No. 11775077, and by the Science and Technology Innovation Plan of Hunan province under Grant No. 2017XK2019.
\end{acknowledgments}

\end{document}